\documentclass[journal=nalefd,manuscript=letter,layout=twocolumn]{achemso}
\setkeys{acs}{keywords = true}
\usepackage{multirow}
\usepackage{adjustbox}
\usepackage{bm}
\usepackage[version=3]{mhchem} 
\usepackage[T1]{fontenc} 
\usepackage{chemformula} 
\usepackage{etoolbox}

\usepackage{blindtext}

\makeatletter
\patchcmd{\acs@contact@details}{E}{*\,E}{}{}
\makeatother

\author{Xirui Gou}
\affiliation{Department of Electrical and Computer Engineering, Carnegie Mellon University, Pittsburgh, PA 15213, USA}
\altaffiliation{These authors contributed equally to this work.}
\author{William Privratsky}
\affiliation{Department of Electrical and Computer Engineering, Carnegie Mellon University, Pittsburgh, PA 15213, USA}
\altaffiliation{These authors contributed equally to this work.}
\author{Wenhan Sun}
\affiliation{Department of Electrical and Computer Engineering, Carnegie Mellon University, Pittsburgh, PA 15213, USA}
\author{Yuncong Liu}
\affiliation{Department of Electrical and Computer Engineering, University of Florida, Gainesville, FL 32611, USA}
\author{Hamed Abiri}
\affiliation{School of Electrical and Computer Engineering, Georgia Institute of Technology, Atlanta, GA 30332 USA}
\author{Qing Li}
\affiliation{Department of Electrical and Computer Engineering, Carnegie Mellon University, Pittsburgh, PA 15213, USA}
\email{qingli2@andrew.cmu.edu}

\title{Chip-scale optically driven phononic frequency comb with 1-70 GHz span}

\abbreviations{FC}

\begin{document}

\begin{abstract}
A phononic frequency comb consists of equally spaced components in the mechanical frequency domain and holds promise for numerous applications. Yet, prior demonstrations have been limited in spectral range due to the inherently low mechanical frequencies. In this work, we report a phononic comb with a record span from 1 to 70 GHz. This result is achieved by harnessing the strong mechanical nonlinearity of a $2.5$-$\mu$m-radius silicon carbide microdisk, which supports a radial breathing mode at $1.655$ GHz with a mechanical quality factor of 13,500. With just 1 mW of dropped optical power, radiation pressure from a continuous-wave pump drives strong phonon lasing, generating 42 phase-locked harmonics with $1.655$ GHz spacing. The combination of such broad bandwidth, low phase noise (-132 dBc/Hz at 1 MHz offset frequency) and frequency stability ($<10^{-7}$ at 1 second of averaging time) positions this ultracompact phononic comb as a powerful platform for diverse applications. 
\end{abstract}


\section{Introduction}

\begin{figure*}[ht]
\centering
\includegraphics[width=0.85\linewidth]{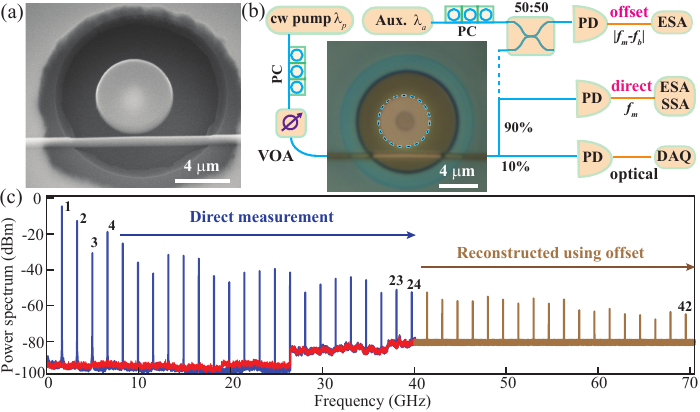}
\caption{(a) Scanning electron micrograph of an undercut $2.5$-$\mu$m-radius silicon carbide microdisk. (b) Experimental schematic. PC: polarization controller; VOA: variable optical attenuator; PD: photodetector; DAQ: data acquisition card; ESA: electrical spectrum analyzer, and SSA: signal source analyzer. (c) Spectrum of the phononic frequency comb consisting of 42 harmonics from the direct ESA measurement (0-40 GHz, blue) and those reconstructed from the heterodyne detection (40-70 GHz, brown). The red line depicts the noise floor of the ESA without input.}
\label{Fig_schematic}
\end{figure*}

A frequency comb is a coherent source made up of equally spaced components in the frequency domain. Its distinctive feature, i.e., the maintenance of phase locking among the components, makes it a powerful tool for applications across science and technology. Optical frequency combs, for instance, have already transformed fields such as optical communications \cite{Moss_comb_communication_review}, precision metrology \cite{Diddams_comb_review1}, imaging \cite{Vahala_comb_imaging}, sensing \cite{Kippenberg_comb_ranging}, microwave photonics \cite{Yi_microwave_nature, Gaeta_microwave_nature, Diddams_microwave_Nature}, and quantum information \cite{Review_quantum_comb}. Much of this progress in the past decade has been driven by the microcomb technology, which leverages optical nonlinearity in low-loss microresonators to significantly reduce device size, weight, and power consumption. In contrast, phononic frequency combs (PFC, also known as mechanical frequency combs)\cite{Seshia_phononiccomb_2017}, which serve as the mechanical analog of optical frequency combs, remain less developed, particularly in terms of demonstrated spectral coverage and phase stability across the spectrum \cite{YangLan_Nature_opto_soliton, Weig_PRX2022_nems_comb,ncomm23_comb_overtone, Dong_optocomb_2024PRL, nanolett2024_soliton_comb, rahmanian_elastomeric_2025, Lei_selfinjection_mechnical}.

The free spectral range (or repetition rate) of a PFC is determined by the physical mechanism driving its formation. When nonlinear energy transfer occurs between two or more mechanical modes with different eigenfrequencies, the repetition rate is generally smaller than the eigenfrequency itself, and the resulting comb can be viewed as modulated sidebands of the excited mode(s) \cite{Seshia_phononiccomb_2017,YangLan_Nature_opto_soliton,Dong_optocomb_2024PRL,nanolett2024_soliton_comb,rahmanian_elastomeric_2025,Lei_selfinjection_mechnical}. Alternatively, overtone frequency combs can emerge when the repetition rate matches the eigenfrequency of a mechanical mode, in which case the spectral components correspond directly to its harmonic overtones \cite{Wong_phononcomb_2014,ncomm23_comb_overtone}.

\section{Results and Discussion}
In this work, we demonstrate an overtone PFC in a compact silicon carbide (SiC) optomechanical resonator with an integrated optical waveguide (Fig.~1). The device, shown in Fig.~1(a), is a $2.5$-$\mu$m-radius microdisk fabricated on a 4H-SiC-on-insulator (4H-SiCOI) chip consisting of a 600-nm 4H-SiC layer on 2-$\mu$m silicon dioxide \cite{Li_4HSiC_comb}. The pattern is defined by electron beam lithography (EBL) and transferred to the SiC layer using CHF$_3$/O$_2$ plasma etching. A second EBL step opens an undercut window, followed by wet oxide etching to release the disk\cite{Li_4HSiC_undercutdisk}. Interestingly, the mechanical quality factor ($Q_m$) of the $2.5$-$\mu$m-radius disk exhibits a local maximum at an undercut ratio (undercut width normalized by the radius) around $(55-60)\ \%$ (see Fig.~1(b)), while further undercutting degrades $Q_m$ until the undercut ratio is above $90\%$ \cite{Tang_Si_disk}. This behavior is qualitatively different than what is reported in the literature, where a large undercut ratio is typically required for obtaining high $Q_m$s \cite{Barclay_diamond_disk, Li_4HSiC_undercutdisk}. A detailed study of anchor-loss behavior in such small 4H-SiC microdisks is left for future work. 

\begin{figure*}[ht]
\centering
\includegraphics[width=0.95\linewidth]{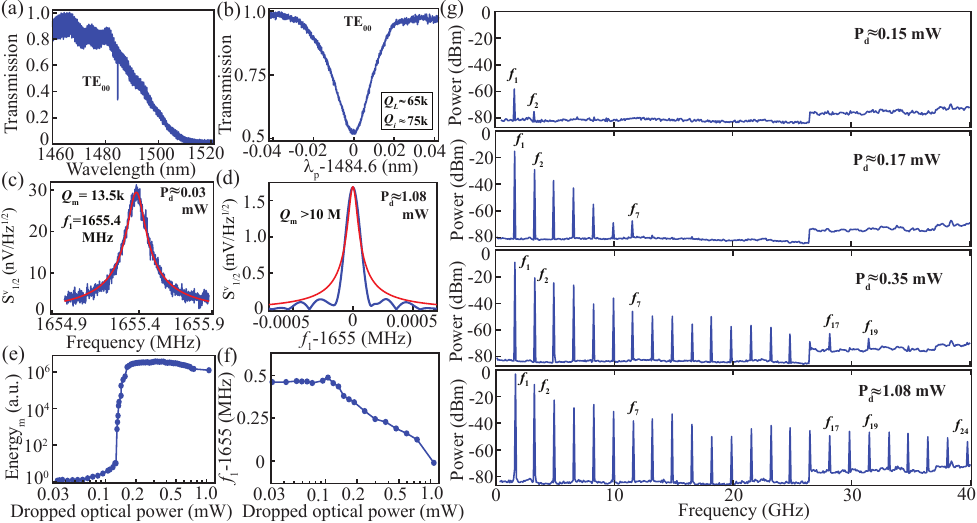}
\caption{(a) Linear optical transmission of the $2.5$-$\mu$m-radius SiC microdisk, in which only the fundamental transverse-electric (TE$_{00}$) resonance around 1485 nm is observed. (b) Zoom-in scan of the TE$_{00}$ resonance in (a). (c) Measured spectral density in voltage (blue solid line) and its Lorentzian fitting (red solid line) of the fundamental radial breathing mode (RBM) of the SiC disk corresponding to a dropped optical power of 30 $\mu$W (red line represents fitting). The resolution bandwidth of the ESA is set at 20 Hz. (d) Spectral density in voltage captured within 10 ms (blue solid line) and its Lorentzian fitting (red solid line) for a dropped optical power of $1.08$ mW with the resolution bandwidth set at 100 Hz. The instantaneous linewidth is estimated to be less than 150 Hz (corresponding $Q_m$ above 10 million). (e) and (f) depict the mechanical energy and resonance frequency of the fundamental RBM as a function of the dropped optical power, respectively. (g) Full-range ESA spectrum corresponding to various dropped optical powers.}
\label{Fig_OEM}
\end{figure*}

The experimental setup for optical and mechanical characterization of the SiC optomechanical resonator is shown in Fig.~1(b). For optical measurements, a continuous-wave tunable laser (Toptica CTL1500, 1460-1580 nm) is used to sweep the resonances of the microdisk, with its transmission detected by a narrow-bandwidth photodetector with high voltage gains (Thorlabs PDB450C). Once an optical mode is identified, the laser detuning and power are adjusted for mechanical mode characterization and PFC generation. For example, for linear mechanical response measurements, a 12-GHz-bandwidth photodetector (Newport AD-40) is combined with a real-time spectrum analyzer (Tektronix RSA5106A, $6.2$-GHz bandwidth), while the optical power is kept sufficiently low to avoid back-action. For broadband PFC spectra, the detection is switched to a 33-GHz-bandwidth photodetector (Finisar XPRV2022A) and a broadband ESA (Rohde $\&$ Schwarz FSEK30, 20 Hz-40 GHz). To access overtones beyond 40 GHz, a second tunable laser (Agilent 81680) with a frequency offset from the pump is introduced, enabling heterodyne down-conversion of higher-order tones into the measurable range of the ESA (details in Fig.~3). Figure 1(c) presents the PFC spectrum obtained directly from the ESA (0-40 GHz) together with the heterodyne-reconstructed spectrum extending up to 70 GHz, with further details discussed below.

In the optical characterization, light is coupled between the input/output fibers and the on-chip waveguide via grating couplers, with an approximate total insertion loss of 19 dB. Owing to the small disk size, the swept-wavelength scan typically reveals only one or two resonances in the scan range ($2.5$-$\mu$m radius corresponds to an optical free spectral range near 7 THz). The access waveguide geometry (600-nm width, 400-nm gap) is optimized to preferentially excite the fundamental transverse-electric (TE$_{00}$) mode, which exhibits the highest optical $Q$. As shown in Figs.~2(a) and 2(b), the measured intrinsic and loaded $Q$ factors of the TE$_{00}$ resonance are 75,000 and 65,000, respectively, near 1485 nm. The optical power dropped into the microdisk ($P_d$) is estimated as $P_d\approx P_{in}(1-T_o)$, with $P_{in}$ and $T_o$ representing the optical power in the waveguide and the normalized optical transmission, respectively.

The fundamental radial breathing mode (RBM) of the $2.5$-$\mu$m-radius microdisk is readily observed in the ESA spectrum with dropped optical power as low as 30 $\mu$W (Fig.~2(c)). The mode exhibits a resonance frequency of $1.655$ GHz and $Q_m$ of 13,500, yielding an $f$-$Q_m$ product of 22 THz-among the highest reported for undercut microdisks across integrated photonic platforms \cite{Barclay_diamond_disk,Li_4HSiC_undercutdisk}. When the dropped power exceeds $120$ $\mu$W, strong phonon lasing is observed, accompanied by a rapid increase in mechanical energy (Fig.~2(e)) and a slight frequency downshift due to the optical spring effect \cite{Barclay_diamond_disk} (Fig.~2(f)). At the highest dropped optical power ($\approx$ $1.08$ mW), the RBM spectrum (Fig.~1(d)) reveals a full width at half maximum below 150 Hz, corresponding to $Q_m>10^7$. The measurement uncertainty is primarily due to rapid frequency fluctuations of several hundred Hz on sub-second timescales (details available in Fig.~4(a)). To mitigate this, the spectrum in Fig.~2(d) was acquired with a real-time ESA (Tektronix), enabling millisecond-scale scans with a relatively small resolution bandwidth (100 Hz). In Fig.~2(g), the full-range ESA spectra (0-40 GHz) at different optical powers portray the evolution of the PFC, with the number of observable harmonics increasing from 2 at $0.15$ mW to 24 overtones at $1.08$ mW. It is also noting that no subharmonics are observed in the spectrum \cite{Wong_phononcomb_2014}. 
\begin{figure}[ht]
\centering
\includegraphics[width=1.0\linewidth]{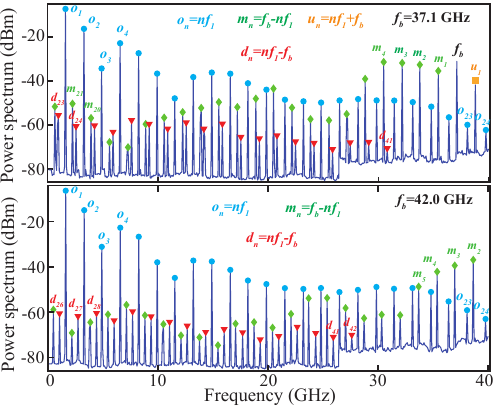}
\caption{Representative ESA spectra in the heterodyne detection scheme (see Fig.~1(b)): the top and bottom panels correspond to an offset frequency (the frequency difference between the pump laser and the auxiliary laser) of $37.1$ and $42.0$ GHz, respectively. Here the circles, diamond, and downward triangles are to indicate the spectral positions of the original ($o_n$), mirrored ($m_n$), and down-shifted ($d_n$) tones of the harmonics. The top panel includes one additional up-shifted ($u_n$, sqaure) tone in the spectrum.}
\label{Fig_beatnote}
\end{figure}

Additional comb lines beyond the ESA’s 40-GHz limit are accessed by mixing the transmitted signal with an auxiliary tunable laser offset by frequency $f_b$ using a 3-dB coupler(see Fig.~1(b)). This scheme resembles heterodyne detection in coherent optical communications, except that the auxiliary laser (serving as the local oscillator) is free-running and not phase-locked to the pump. As a result, the detected RF tones associated with $f_b$ experience power fluctuations arising from time-varying phase. Their power is estimated by recording the maximum values from multiple ($>10$) fast ESA scans with varied offset frequencies. Two representative cases are shown in Fig. 3: in the top panel ($f_b\approx 37.1$ GHz), the spectrum contains the original harmonic tones ($O_n$), the beatnote signal of the two lasers ($f_b$), mirrored tones ($m_n$), and down-shifted tones ($d_n$). For the bottom panel, the beatnote tone $f_b$ ($\approx 42$ GHz) is out of the ESA range, but similar mirrored and down-shifted tones are still observed. 

\begin{figure}[ht]
\centering
\includegraphics[width=1.0\linewidth]{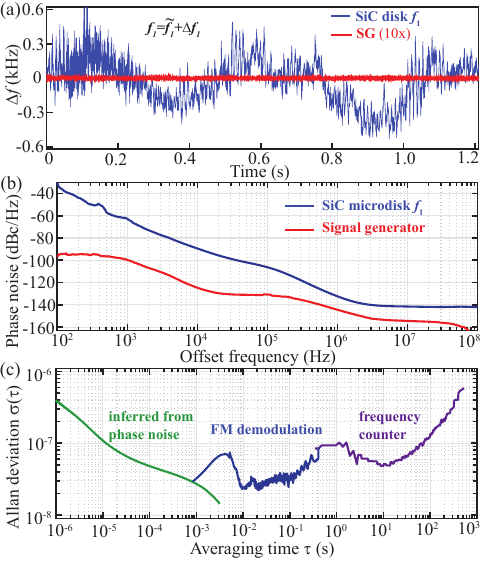}
\caption{(a) Real-time frequency fluctuation ($\Delta f_1$) of the fundamental tone around its nominal value ($\tilde f_1$) from the SiC microdisk (blue line) compared against that from a signal generator tuned to the same frequency (SG, red line, multiplied by a factor of 10 for visibility).(b) Single-sideband phase noise of the fundamental tone (blue solid line) and the output from a signal generator (red line). (c) Allan deviation: the green line is inferred from the phase noise in (b); the blue line is calculated based on the demodulated frequency fluctuation in (a); and the purple line is acquired using the frequency counter in ESA.}
\label{Fig_PN}
\end{figure}

For microwave applications, the phase noise and frequency stability of the generated PFC are critical performance metrics. We first examine the fundamental tone at $1.655$ GHz under maximum optical power. Using the ESA's analog frequency demodulator, we track its real-time frequency fluctuations around the nominal value, which are on the order of $100$ Hz over $10$ ms timescales and can increase to $\sim 1$ kHz within $1$ s (Fig.~4(a)). These fluctuations likely arise from multiple sources, including the free-running pump laser (whose detuning is not locked to the cavity) and variations in on-chip power due to laser and fiber-chip coupling instabilities. Despite these imperfections, the single-sideband phase noise measured with a signal source analyzer (Agilent E5052B) reaches $-90$ dBc/Hz, $-106$ dBc/Hz, and $-132$ dBc/Hz at $10$ kHz, $100$ kHz, and $1$ MHz offsets, respectively (Fig.~4(b)). Compared to a state-of-the-art signal generator (Rohde $\&$ Schwarz SMU200A) at similar frequencies, the SiC microdisk exhibits noticeably higher phase noise below $10$ kHz, consistent with the frequency fluctuations observed in Fig.~4(a). The Allan deviation (Fig.~4(c)) is derived using different methods depending on the averaging timescale: for averaging times below $1$ ms, it is mainly estimated from the measured phase noise (offsets $>1$ kHz) in Fig.~4(b); for averaging times between $1$ ms and $0.3$ s, it is calculated from the demodulated frequency fluctuations in Fig.~4(a); and for averaging times longer than $0.3$ s, direct frequency tracking is performed using the frequency counter in ESA. Overall, the oscillator achieves a frequency stability better than $10^{-7}$ at 1s of averaging time, with degradation at longer times attributed to environmental temperature fluctuations.

\begin{figure}[ht]
\centering
\includegraphics[width=1.0\linewidth]{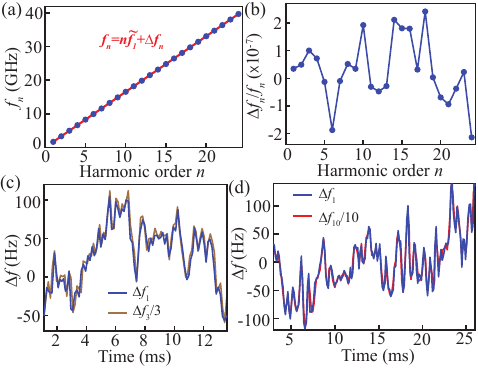}
\caption{(a) Measured center frequencies of harmonic tones (blue circles) and their linear fitting (red line). (b) Fractional residual frequency errors for each harmonic from the linear fitting. (c) Simultaneously detected frequency fluctuation of the fundamental tone (blue) compared against that of the third harmonic (brown, amplitude divided by three). (d) Similar to (c) but for comparison between the fundamental (blue) and the 10-th harmonic (red).}
\label{Fig_locking}
\end{figure}

Next, we examine the unique phase-locking property among the spectral components of our PFC. As shown in Fig.~5(a), frequency measurements using ESA confirm a linear relationship between the overtone frequency ($f_n$) and the harmonic order ($n$), given by $f_n = n\tilde{f_1} + \Delta f_n$, where $\tilde{f_1}$ is the nominal fundamental frequency and $\Delta f_n$ is the residual fitting error. Each $\Delta f_n$ fluctuates with time, leading to a fractional uncertainty $\Delta f_n / f_n$ on the order of $10^{-7}$ (Fig.~5(b)). If the overtone FC exhibits true phase locking with equal spacing, the relation $f_n = n f_1$ should hold exactly, even as $f_1$ undergoes random fluctuations (see Fig.~4(a)). As such, the fitting errors representing real-time frequency fluctuations should scale as $\Delta f_n = n \Delta f_1$. To verify this, we employ analog frequency demodulation using two synchronized ESAs: one (Tektronix RSA5106A) for the fundamental tone and the other (FSEK30) for the overtone. Representative measurements for the third and tenth harmonics, shown in Fig.~5(c) and 5(d), respectively, confirm strong phase locking across the comb lines. This phase-locking behavior is a defining feature of phononic frequency combs and underpins their potential as coherent, low-noise microwave sources.

\section{Conclusion}
We demonstrated an ultracompact phononic frequency comb based on a $2.5$-$\mu$m-radius silicon carbide optomechanical resonator, capable of generating 42 phase-locked harmonics with a frequency spacing of $1.655$ GHz under a dropped optical power of only $\sim$1 mW. The resulting comb spans a record 1–70 GHz and exhibits low phase noise (–132 dBc/Hz at a 1 MHz offset for the fundamental tone) and excellent frequency stability ($<10^{-7}$ at 1s of averaging time). These performance metrics can be further improved through techniques such as laser injection locking of the SiC microdisk resonator and active feedback for compensating environmental temperature drifts. With its chip-scale form factor, broadband coverage, and phase coherence, this SiC-based phononic frequency comb holds strong potential for applications ranging from high-performance microwave sources in 5G/6G communications to quantum information technologies involving high-frequency microwave photons.

\section{Funding}
This work was supported by NSF (2240420, 2131162, 2427228) , CableLabs University Outreach Program, and 2025 PQI Community Collaboration Award.  

\begin{acknowledgement}
The authors acknowledge helpful discussions on the phase noise measurement with Prof.~Gary Fedder and Prof.~Gianluca Piazza at CMU, the use of Bertucci Nanotechnology Laboratory at Carnegie Mellon University supported by grant BNL-78657879, and the Materials Characterization Facility supported by grant MCF-677785. 
\end{acknowledgement}


\section{Disclosures}  The authors declare no conflicts of interest.

\section{Data Availability} Data underlying the results presented in this paper are not publicly available at this time but may be obtained from the authors upon reasonable request.

\bibliography{SiC_Ref}

\end{document}